# Controlling the Sense of Molecular Rotation

Sharly Fleischer, Yuri Khodorkovsky, Yehiam Prior  and
Ilya Sh. Averbukh[1]

*Department of Chemical Physics, Weizmann Institute of Science, Rehovot 76100, Israel*
*ilya.averbukh@weizmann.ac.il, phone +972-8-934-3350, fax +972-8-934-4123*

**Abstract:** We introduce a new scheme for controlling the sense of molecular rotation. By varying the polarization and the delay between two ultrashort laser pulses, we induce unidirectional molecular rotation, thereby forcing the molecules to rotate clockwise/counter-clockwise under field-free conditions.  We show that unidirectionally rotating molecules are confined to the plane defined by the two polarization vectors of the pulses, which leads to a permanent anisotropy in the molecular angular distribution. The latter may be useful for controlling collisional cross-sections and optical and kinetic processes in molecular gases. We discuss the application of this control scheme to individual components within a molecular mixture in a selective manner.

The essence of coherent control is to drive a molecular system towards specific behavioural goal. The goals are usually set as the enhanced population of a specific vibrational or electronic state and the tools are ultrashort laser pulses which are modulated either in the time or frequency domains. Here we present a double pulse scheme for controlling the sense (clockwise / counter clockwise) of the molecular rotation.

Laser induced molecular rotation and alignment has received significant attention in recent years. In the last decade, interest in the field has increased, mainly due to the improving capabilities to control the laser pulse characteristics (such as time duration and temporal shape), which in turn leads to potential applications offered by controlling the angular distribution of molecules.   Since the typical rotational motion is 'slow' (~10 ps) with respect to the typical short pulse (~50 fs), effective rotational control and manipulation are in reach. In the liquid phase, molecular alignment following excitation by a strong laser pulse was observed in the seventies[1], and proposed as a tool for optical gating. In the early experiments, picosecond laser pulses were used for the excitation, and deviation of the refractive index from that of an isotropic gas was utilized as a measure of the alignment[2,3]. More recently, this research area has been revisited both theoretically and experimentally (for a recent review, see Ref.[4]). Temporal rotational dynamics of pulse-excited molecules was studied[5,6,7], and multiple pulse sequences giving rise to the enhanced alignment were suggested,[8,9] and realized[10,11,12,13]. Further manipulations such as optical molecular centrifuge and alignment-dependent strong field ionization of molecules were demonstrated[13,14,15]. Selective rotational excitation in bimolecular mixtures was suggested and demonstrated in the mixtures of molecular isotopes[16] and spin isomers[17]. Transient molecular alignment has been shown to compress ultrashort light pulses[18,19] and it is successfully used in controlling high harmonic generation[20,21,22]. Other experiments were reported in which transient grating techniques were employed for detailed studies of molecular alignment and deformation[23,24]. In the past few years, molecular alignment became a common tool in attosecond studies, in particular, in experiments for probing molecular bond structures[25,26].

---

[1] *also at Kavli Institute for Theoretical Physics, University of California, Santa Barbara, CA 93106,  USA*

In practically all the previous works in the field of laser molecular alignment (with the exception of Ref.[14]), the rotational motion was enhanced, but the net total angular momentum delivered to the molecules remained zero, and for a good reason. For single pulse schemes, as well as for techniques using multiple pulses polarized in the same direction, no preferred sense of rotation exists due to the axial symmetry of excitation.

In order to inject angular momentum to the medium and to force the molecules to rotate with a preferred sense of rotation, one has to break the axial symmetry. This has been previously demonstrated by Karczmarek *et. al*[14] who used two oppositely chirped, circularly polarized pulses overlapping in time and space, thereby creating a linearly polarized pulse, rotating unidirectionally and accelerating in a plane.

In solid state, controlled unidirectional rotation of induced polarization by impulsive excitation of two-fold degenerate lattice vibrations was demonstrated in $\alpha$-quartz[27].

In this paper, we propose a double pulse scheme for breaking the axial symmetry and for inducing unidirectional molecular rotation under field-free conditions.

Pictorially, our double pulse control scheme is sketched in Figure 1. An ultrashort laser pulse (red arrow), linearly polarized along the $z$ axis, is applied to the molecular ensemble and induces coherent molecular rotation. The molecules rotate under field-free conditions until they reach an aligned state, in which they are temporarily confined in a narrow cone around the polarization direction of the first pulse. At this moment, a second pulse, linearly polarized at 45 degrees to the first one, is applied, inducing unidirectional (clockwise, in our case) molecular rotation.

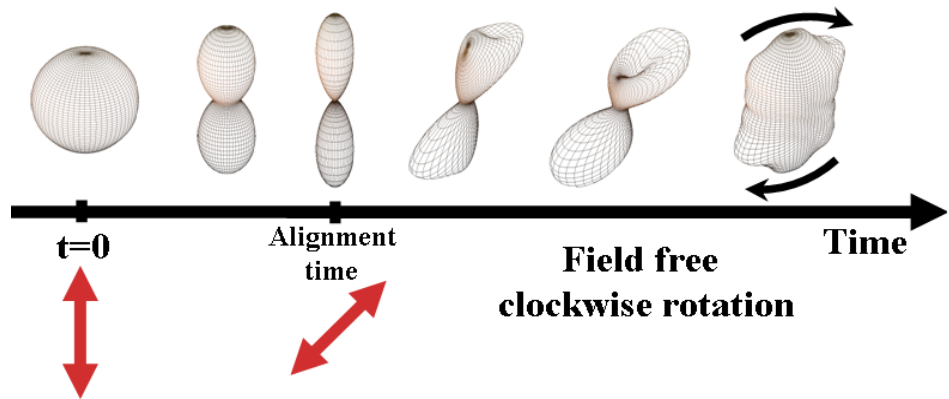

Figure 1. Double-pulse scheme for excitation of unidirectional molecular rotation.

In the present paper, we discuss the dependence of the induced angular momentum on the pulse intensities, and the delay between pulses. Moreover, we focus on the molecular angular distribution when the molecules are subject to unidirectional rotation, and show that it is confined in the plane defined by the two polarization vectors of the pulses. This anisotropic angular distribution is characterized by the observable $\langle \cos^2 \varphi \rangle$ which is referred to as the azimuthal factor. This highly anisotropic angular distribution induced by the breaking of the axial symmetry may offer an efficient way to control the kinetic and optical properties of the gas medium.

We consider the problem by modelling the molecules as driven rigid rotors interacting with a linearly polarized laser field. Within this model the Hamiltonian of the linear molecules is given by

$$H = \frac{\hat{J}^2}{2I} + V(\theta, t), \qquad (1)$$

where $\hat{J}$ is the angular momentum operator, $\theta$ is the angle between the polarization vector of the field (defining the $z$ axis) and the molecular axis, and $I$ is the moment of inertia of the molecule. The latter is related to the molecular rotational constant $I = \hbar/(4\pi cB)$, where $\hbar$ is Planck constant and $c$ is the speed of light. The interaction term is given by

$$V(\theta,t) = -\tfrac{1}{4}\varepsilon^2(t)[(\alpha_{\parallel} - \alpha_{\perp})\cos^2(\theta) + \alpha_{\perp}] \qquad (2)$$

where $\varepsilon(t)$ is the envelope amplitude of the laser field, and $\alpha_{\parallel}$, $\alpha_{\perp}$ are the parallel and perpendicular components of the polarizability tensor, respectively.

We simulated the proposed scheme quantum-mechanically by two independent methods. The first of them is mainly analytical: it uses spectral decomposition of the time-dependent rotational wave function, and relies heavily on angular momentum algebra for calculating the observable quantities (see Appendix A for the mathematical details,). Such an approach has been widely used in the past by many groups, including ourselves [8,9], to analyze multipulse alignment,. The second method uses direct numerical simulation of the driven ensemble of quantum rotors by means of Finite-Difference Time-Domain (FDTD) approach (for details, see Appendix B). In both cases, the laser pulses were approximated as $\delta$ functions, and their integrated 'strength' was characterized by a dimensionless pulse strength parameter

$$P = (\Delta\alpha/4\hbar)\int_{-\infty}^{+\infty}\varepsilon^2(t)dt \qquad (3)$$

where $\Delta\alpha = \alpha_{\parallel} - \alpha_{\perp}$. Physically, the parameter $P$ represents a typical increase of the molecular angular momentum (in the units of $\hbar$) due to the interaction with the pulse. All simulations were performed at finite temperature, and the results were averaged over a thermal molecular ensemble. Moreover, recently the same problem was studied by us classically with the help of the Monte Carlo method, the details are given elsewhere[28]. All three approaches gave qualitatively similar outcome, with two quantum treatments (presented here) yielding essentially identical results. For the sake of brevity, in the body of the paper we present the results of the quantum calculations briefly, and the readers are referred to Appendices A,B where the details of calculations are given.

We consider separately the evolution of molecules starting from individual eigenstates of the rigid rotor (described by a spherical harmonic $Y_l^m(\theta,\varphi)$). The action of the first laser pulse (polarized in $z$ direction) was approximated by impulsive excitation and then the resulting wavepacket was propagated in time using either the spectral decomposition of the wavefunction

$$\psi(\theta,\varphi,t) = \sum_{l} C_{l,m} e^{-iE_l t/\hbar} Y_l^m(\theta,\varphi) \qquad (4)$$

(here $E_l$ is the energy of the $l$-th eigenstate of the rigid rotor), or by means of direct FDTD simulation. Next, we calculated the thermally averaged alignment factor $\langle \cos^2\theta \rangle$ and found the time of maximal alignment, just before $\tfrac{1}{2}T_{rev}$ (see inset to Figure 2). Here $T_{rev} = 1/(2Bc)$ is the rotational revival time. At this time we applied the second $\delta$-pulse, linearly polarized in the $xz$-plane at an angle $\theta_p$ with respect to the vertical $z$-direction. The second pulse creates much richer wavepackets that are composed of states having different $m$ quantum numbers. By using the spectral decomposition again (Appendix A), or FDTD simulation (Appendix B) we calculated the angular momentum and the time-dependent azimuthal factor $\langle \cos^2\varphi \rangle$ after the second pulse. Finally, the results were thermally averaged by repeating the calculation for different initial states $Y_l^m(\theta,\varphi)$, and adding them with proper statistical weights.

In Figure 2, we present the calculated expectation value of the $y$-component of the angular momentum ($<J_y>$) as a function of the polarization angle of the second pulse. The second pulse is applied at the time of maximal alignment (marked by an arrow in the inset) just before the one-half rotational revival following the first aligning pulse.

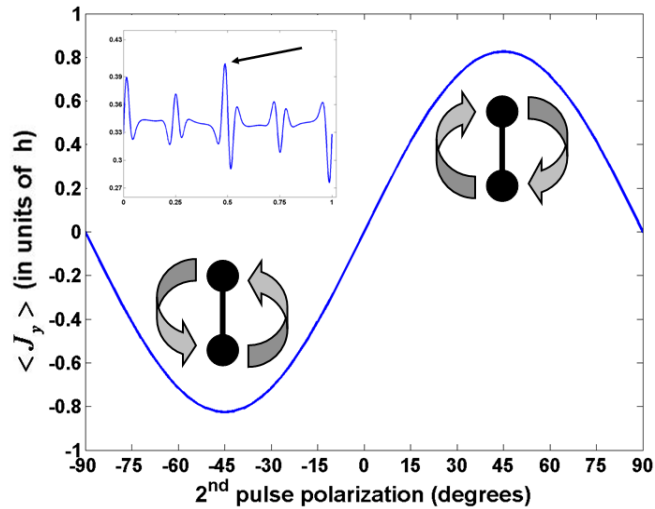

**Figure 2**: Expectation value of the angular momentum along the $y$ axis as a function of the polarization angle of the second pulse. The pulse intensities are given by $P_1 = 3$, $P_2 = 6$ (see text for definitions), the temperature is 100 K (for Nitrogen molecules). The second pulse is applied at the maximally aligned state, just before $\tfrac{1}{2}T_{rev}$. The direction of the molecular rotation is depicted by the cartoons, showing clockwise and counter-clockwise rotation for $45°$ and $-45°$, respectively.

The maximal angular momentum is achieved when the second pulse is polarized at $\pm 45°$ with respect to the first pulse polarization. This result is in agreement with the classical description of the light–rotor interaction. The interaction term in the Hamiltonian is given by $V = -\tfrac{1}{4}\varepsilon^2 \Delta\alpha \cos^2\theta$, therefore the torque applied to a molecule oriented at angle $\theta$ to the pulse polarization is given by $\tau(\theta) \propto -\dfrac{dV}{d\theta}$, and the corresponding angular velocity gained by the molecule is $\omega(\theta) \propto -\sin(2\theta)$. Thus, the maximal velocity is gained by a molecule initially aligned at $\theta = \pm 45°$ to the field. As the next step we explore the dependence of the induced angular momentum on the strength $P_{1,2}$ of the two laser pulses. Figure 3 shows $\langle \cos^2\theta \rangle$ of the aligned state just before $1/2 T_{rev}$ as a function of the first pulse strength $P_1$.

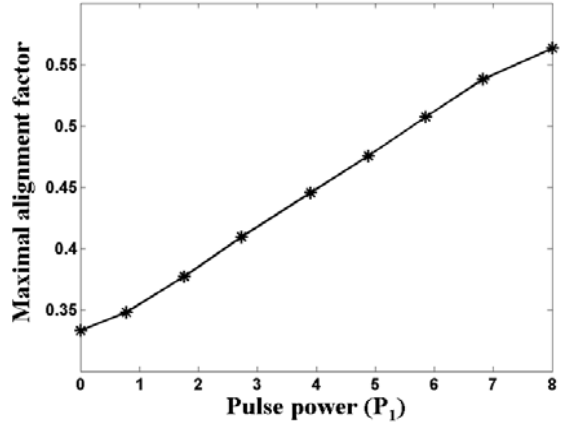

Figure 3: Maximal alignment factor of the aligned state, just before $\tfrac{1}{2}T_{rev}$ for different pulse strength ($P_1$) at 150K (for Nitrogen molecules). The dependence is fairly linear in the shown parameter region. At higher pulse power, the alignment factor saturates at ~ 0.9.

Figure 4 shows the angular momentum induced by the double-pulse excitation. We scan the power of the second pulse ($P_2$) for each of the $P_1$ values shown in Figure 3.

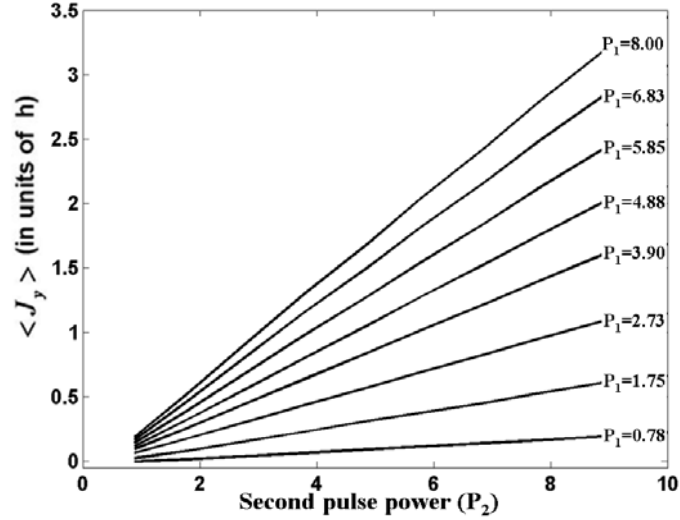

Figure 4: Induced angular momentum as a function of the strength of the second pulse, $P_2$ calculated for different strength $P_1$ of the first pulse at 150K (for Nitrogen molecules).

As $P_1$ and $P_2$ increase, so does the induced angular momentum. Looking at the line slopes in Figure 4, one can immediately deduce a clear trend: as the power of the first pulse increases, so does the slope, i.e. the ability of inducing angular momentum by the second pulse increases with the first pulse power (and the corresponding alignment factor). If the total energy of two laser pulses is fixed ($P_1 + P_2 = P_{tot}$), one may ask what is the best pair of pulses leading to the maximal induced angular momentum. Figure 5 depicts $<J_y>$ as a function of $P_1 - P_2$ for $P_{tot} = 13$ (for rotational temperature of 150K in the case of Nitrogen).

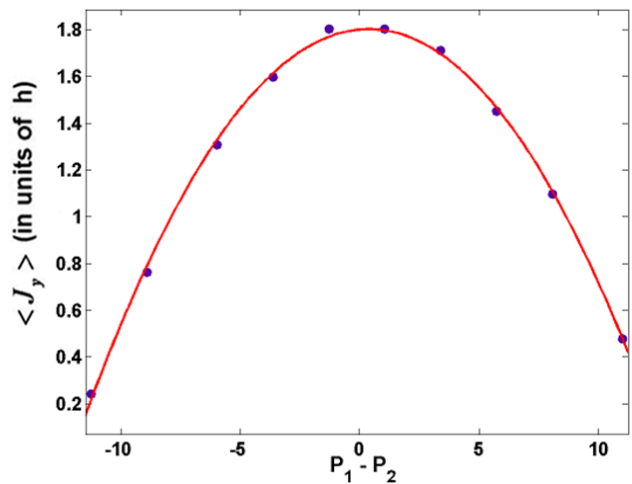

Figure 5: Calculation of the induced angular momentum $\langle J_y \rangle$ as a function of $P_1 - P_2$. The total strength of the two pulses was kept constant $P_{tot} = P_1 + P_2 = 13$, at 150K (for Nitrogen molecules).

Figure 5 shows quadratic dependence of $\langle J_y \rangle$ on $P_1 - P_2$, for a fixed pulse power $P_{tot}$. The maximal angular momentum is found for $P_1 = P_2 = P_{tot}/2$. This result is in agreement with the linear character of the plots shown at Figs. 3 and 4.

Up to now, we have shown that if the second pulse is applied at the time of maximal molecular alignment caused by the first pulse (see inset of Figure 2), then the maximal unidirectionality is achieved when the second pulse is polarized $45^0$ to the first one. . In what follows, we keep the polarization at $45^0$ and vary the delay between the two laser pulses.

Figure 6 shows the induced angular momentum $\langle J_y \rangle$ for the fixed pulse strengths $P_1$ and $P_2$ as a function of the time delay between the two pulses around $\frac{1}{2}T_{rev}$.

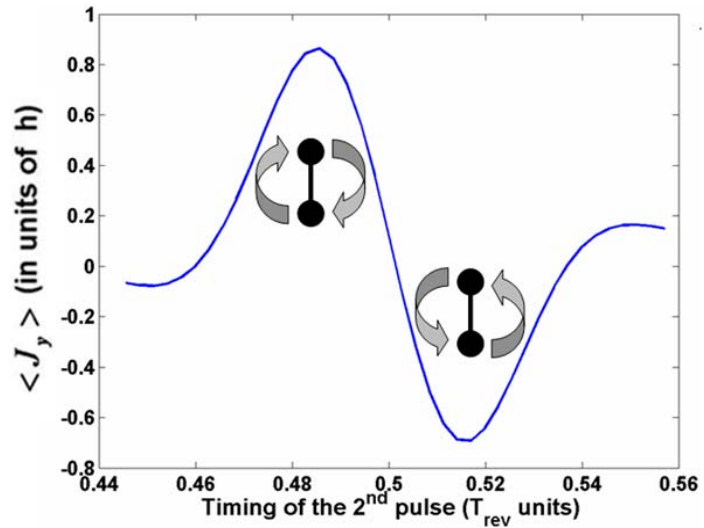

Figure 6. Calculation of the induced angular momentum $\langle J_y \rangle$ as a function of time delay between the pulses around $1/2T_{rev}$, $P_1 = 3$, $P_2 = 6$, the temperature is 100 K (for Nitrogen molecules).

If the second pulse is applied just before the $\frac{1}{2}T_{rev}$, when the molecular distribution peaks in the $z$ direction, the clockwise unidirectional rotation is induced in the molecular medium. But if the second pulse is applied just after $\frac{1}{2}T_{rev}$, when the molecules are anti-aligned, the unidirectional rotation is counter-clockwise. For Nitrogen molecules, the difference between these two time delays is about 200 fs.

In a mixture of two species such as molecular isotopes[16] one can find time delays when one species is aligned while the other is anti-aligned at the same time.

Application of the second pulse at this time moment will result in the opposite senses of rotation induced for the two species, which can be potentially used for their physical separation.

In Figure 7 we show a calculation similar to the one in Figure 6 but now we scan the time delay between the two pulses around $\tfrac{1}{4}T_{rev}$. As we showed recently[17], different spin isomers of homonuclear diatomic molecules experience opposite alignment dynamics in this region. When the second pulse is applied exactly at $\tfrac{1}{4}T_{rev}$, para and ortho nuclear spin isomers acquire opposite senses of rotation. The reason for this phenomenon is the drastically different angular distributions (alignment vs. anti-alignment) that the para and the ortho molecules attain in this time domain[17].

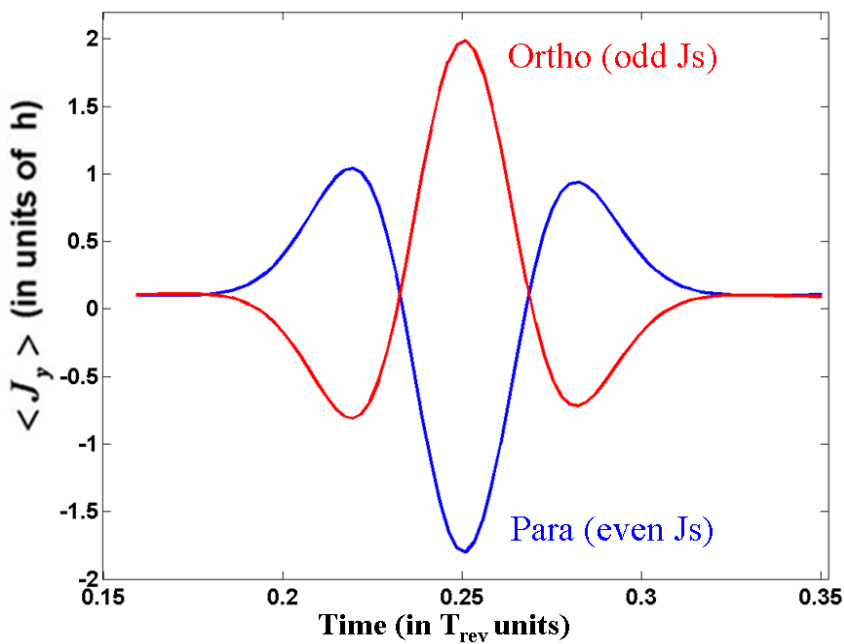

Figure 7: Calculation of the induced angular momentum $\langle J_y \rangle$ as a function of time delay between the pulses around $\tfrac{1}{4}T_{rev}$. $P_1 = 3$, $P_2 = 6$, the temperature is 100 K (for Nitrogen molecules)

Finally, we concentrate on the angular distribution of molecules excited by our double pulse scheme. Classically, we expect that unidirectionally rotating molecules remain confined to the plane defined by the two polarization vectors of the pulses. To investigate this problem in detail, we consider a new observable $\langle \cos^2 \varphi \rangle$ correlated with molecular confinement to this plane (where $\varphi$ is the azimuthal angle). In Figure

8, we plot $\langle \cos^2 \varphi \rangle$ as a function of time. We start from the point where the second pulse is applied, since before that $\langle \cos^2 \varphi \rangle = 1/2$, and it is time independent.

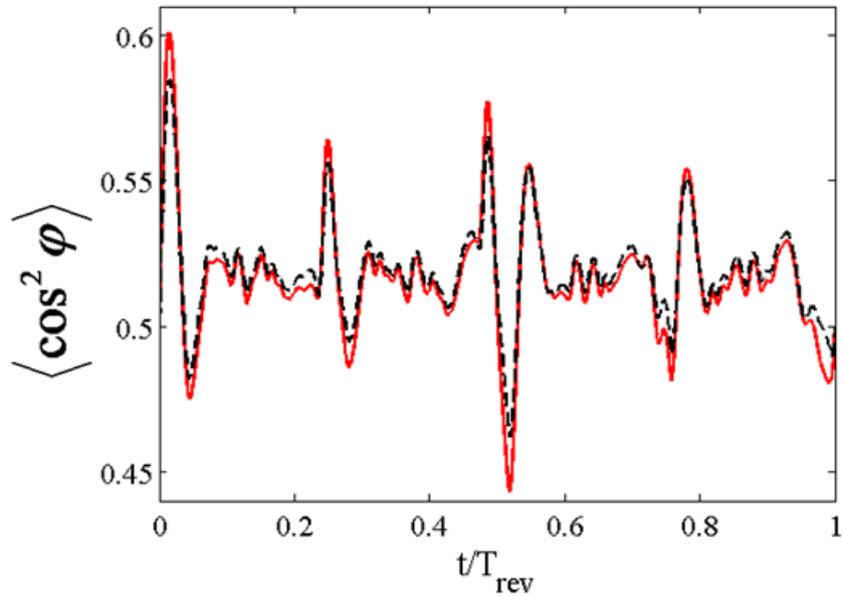

Figure 8: $\langle \cos^2 \varphi \rangle$ as a function of time after the second pulse calculated by the analytical method (dashed black line), and FTDT approach (solid red line). $P_1 = P_2 = 5$, For Nitrogen molecules at temperature of 50K.

One can clearly observe the features of the revival phenomenon manifested in the peaks (dips) of $\langle \cos^2 \varphi \rangle$ at different fractional times within one revival period. Note that in contrast to the alignment factor ($\langle \cos^2 \theta \rangle$), showing only full, half and quarter revivals, the azimutal factor $\langle \cos^2 \varphi \rangle$ clearly exhibits higher fractional revivals.

At higher temperatures and higher excitation powers, the fractional revivals are better observed. In figure 9 we plot $\langle \cos^2 \varphi \rangle$ as a function of time for $P_1 = P_2 = 10$ at 150K for Nitrogen molecules.

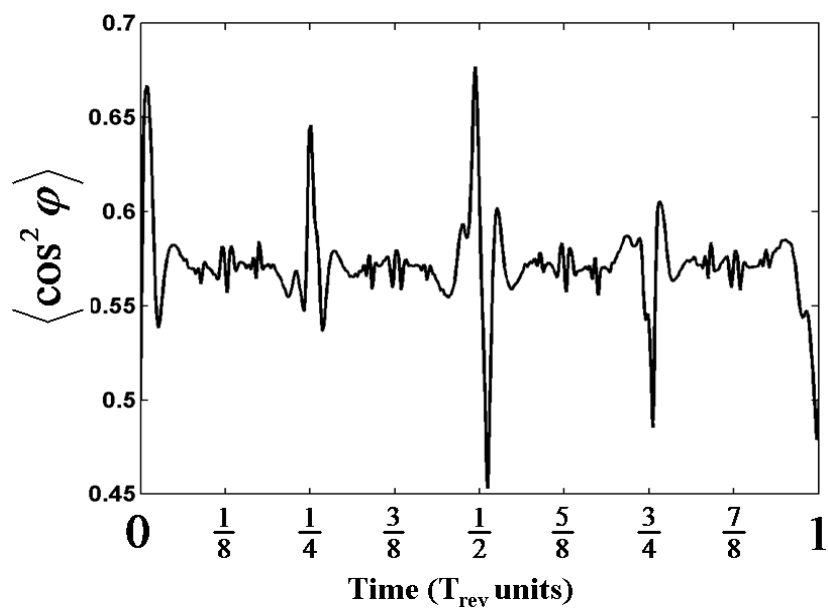

Figure 9: $\langle \cos^2 \varphi \rangle$ as a function of time after the second pulse. $P_1 = P_2 = 10$, 150K (Nitrogen molecules).

Note that fractional revivals around 1/3, 1/6, 1/8 $T_{rev}$ of the revival are clearly observed. Furthermore, the time average value of $\langle \cos^2 \varphi \rangle$ in Fig. 9 is ~0.57, which certainly exceeds the isotropic value of 0.5. This means that the molecular axis indeed preferentially occupies the plane defined by the two polarization directions. For details of the calculation of $\langle \cos^2 \varphi \rangle$ we refer the reader to the Appendices A and B.

Figure 10 depicts the angular distribution (averaged over one full revival period) after the second pulse, confirming the confinement of the molecules to the plane.

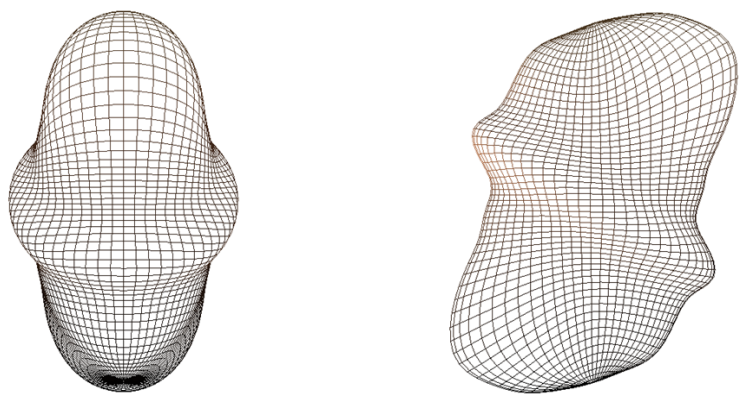

Figure 10: Angular distribution of the unidirectionally rotating molecules averaged over one full revival shown from two perpendicular view directions.

The induced anisotropy leads to anisotropic cross sections for collisions of the molecules between themselves, or with atoms/ molecules of a buffer gas. This may result in controlled anisotropic diffusion of different species in a molecular mixture. Moreover, the persistent anisotropy allows for controlling the gas-surface scattering phenomena. These and other related problems are a subject of an ongoing research.

Summarizing, we have shown that by applying two non-parallel-polarized ultrashort laser pulses one can break the axial symmetry of molecular rotation and control the sense of the rotation, thereby injecting angular momentum to the molecular ensemble. The sense of rotation depends on the relative timing and angle between the polarization directions of the pulses. This double excitation scheme can be used for selective-excitation of unidirectional rotation in mixtures of molecular isotopes and spin isomers. An important outcome is the anisotropic confinement of the molecules to the plane defined by the two polarizations, which leads to anisotropic collisional cross sections which may offer a novel way for controlling kinetic processes in molecular gases.

This research was supported in part by a Grant No. 1186/05 from the Israel Science Foundation, and by the National Science Foundation under Grant No. PHY05-51164. It is made possible in part by the historic generosity of the Harold Perlman Family. SF is a recipient of the Eshkol Scholarship from the Israel Ministry of Science, Culture & Sport. IA is the incumbent of the Patricia Elman Bildner Professorial Chair and YP is the incumbent of the Sherman Professorial Chair in Physical Chemistry.

# Appendix A: Double-pulse scheme: analytical treatment

We consider the action of the laser pulses in the impulsive approximation (i.e. assuming that they are much shorter than the typical time scales of the rotational motion) by treating them as $\delta$- pulses. During the $\delta$-pulse, one may neglect the kinetic energy term in the Schrödinger equation, which can be solved exactly after that. As the result, the rotational wavefuction just after the pulse, $\psi^{(+)}$ is related to the wavefunction $\psi^{(-)}$ just before the pulse by

$$\psi^{(+)} = \exp[iP\cos^2\theta]\psi^{(-)}. \tag{5}$$

Here $P = (\Delta\alpha/4\hbar)\int_{-\infty}^{+\infty}\varepsilon^2(t)dt$ is the dimensionless strength of the pulse. In this calculation the initial states $\psi^{(-)}$ are the eigenstates $Y_l^m(\theta,\varphi)$ of the rigid rotor. To consider the post-pulse evolution, we decompose the resulting $\psi^{(+)}$ in the basis of spherical harmonics. Despite the simple look of the impulsive transformation, this decomposition is a non-trivial task that can be done by several methods (see, e.g.[8,29]). In our case, we introduce the artificial "time"-parameter $\tau$, and consider the $\tau$-dependent construction

$$e^{iP\cos^2\theta\tau}\psi^{(-)} = \sum_{l,m}C_{l,m}(\tau)Y_l^m(\theta,\varphi) \tag{6}$$

We convert the problem to the solution of a set of coupled differential equations for the coefficients $C_{l,m}(\tau)$. At $\tau=0$ only the coefficient corresponding to the initial eigenstate is non-zero (and equal to 1). The after-pulse wavefunction $\psi^{(+)}$ is given by (6) at $\tau=1$. The needed set of differential equations is produced by differentiating (6) with respect to $\tau$, and projecting both sides of the resulting equation to $\langle Y_{l'}^{m'}|$:

$$\dot{C}_{l',m'}(\tau) = iP\sum_{l,m}C_{l,m}(\tau)\langle Y_{l'}^{m'}|\cos^2\theta|Y_l^m\rangle \tag{7}$$

We write $\cos^2\theta$ in terms of the spherical harmonic functions:

$$Y_2^0(\theta,\varphi) = \frac{1}{4}\sqrt{\frac{5}{\pi}}(3\cos^2\theta-1) \Rightarrow \cos^2\theta = \frac{4}{3}\sqrt{\frac{\pi}{5}}Y_2^0(\theta,\varphi) + \frac{1}{3}$$

For the integration of the product of three spherical harmonic functions, we use the Wigner 3j symbol:

$$\int_0^{2\pi}\int_0^{\pi} Y_{l_1}^{m_1} Y_{l_2}^{m_2} Y_{l_3}^{m_3} \sin\theta \, d\theta \, d\varphi = \sqrt{\frac{(2l_1+1)(2l_2+1)(2l_3+1)}{4\pi}} \begin{pmatrix} l_1 & l_2 & l_3 \\ 0 & 0 & 0 \end{pmatrix} \begin{pmatrix} l_1 & l_2 & l_3 \\ m_1 & m_2 & m_3 \end{pmatrix} \quad (8)$$

The equations (7) become:

$$\dot{C}_{l',m'}(\tau) = iP \frac{4}{3}\sqrt{\frac{\pi}{5}} \sum_{l,m} C_{l,m}(\tau) \langle Y_{l'}^{m'} | Y_2^0 | Y_l^m \rangle + \frac{iP}{3} C_{l',m'}(\tau) \quad (9)$$

From the equations above, one can see that the interaction term ($\propto \cos^2\theta$) couples the $|l,m\rangle$ states only to the $|l\pm 2, m\rangle$ states. Thus the resulting wavepacket consists of states having the same parity (odd/even) as the initial state, and with the same $m$ number.

After decomposing $\psi^{(+)}$ in the form of (6) (with $\tau = 1$), we propagate this wavepacket in time for every possible initial state, and calculate the thermally averaged alignment factor $\langle \cos^2\theta \rangle$ as a function of time. At the moment of the maximal alignment, just before $\frac{1}{2}T_{rev}$, we apply the second pulse, polarized in the $xz$ plane at an angle $\theta_p$ with respect to the $z$ axis.

The interaction with the second pulse is proportional to $\cos^2\beta$, where $\beta$ is the angle between the molecular axis and the inclined polarization vector:

$$\cos^2\beta = \cos^2\varphi \sin^2\theta \sin^2\theta_p + \cos^2\theta \cos^2\theta_p + \cos\varphi \sin\theta \cos\theta \sin(2\theta_p) \quad (10)$$

After some algebra we express the angular-dependent functions in (10) as:

$$\cos^2\varphi \sin^2\theta = \sqrt{\frac{\pi}{15}}(Y_2^2 + Y_2^{-2}) - \frac{2}{3}\sqrt{\frac{\pi}{5}} Y_2^0 + \frac{1}{3}$$

$$\cos^2\theta = \frac{4}{3}\sqrt{\frac{\pi}{5}} Y_2^0 + \frac{1}{3} \quad (11)$$

$$\cos\varphi \sin\theta \cos\theta = \sqrt{\frac{2\pi}{15}}(Y_2^{-1} - Y_2^1)$$

To consider the action of the second pulse in the impulsive approximation (similar to (6), (7)) we have to solve the following set of coupled differential equations:

$$\dot{C}_{l',m'}(\tau) = iP \sum_{l,m} C_{l,m}(\tau) \left\{ \begin{array}{l} \sin^2\theta_p \left[ \sqrt{\frac{\pi}{15}} \langle Y_{l'}^{m'} | Y_2^{-2} | Y_l^m \rangle + \sqrt{\frac{\pi}{15}} \langle Y_{l'}^{m'} | Y_2^2 | Y_l^m \rangle + \frac{1}{3} \langle Y_{l'}^{m'} | Y_l^m \rangle - \frac{2}{3}\sqrt{\frac{\pi}{5}} \langle Y_{l'}^{m'} | Y_2^0 | Y_l^m \rangle \right] + \\ +\cos^2\theta_p \left[ \frac{4}{3}\sqrt{\frac{\pi}{5}} \langle Y_{l'}^{m'} | Y_2^0 | Y_l^m \rangle + \frac{1}{3} \langle Y_{l'}^{m'} | Y_l^m \rangle \right] + \\ +\sin(2\theta_p) \left[ \sqrt{\frac{2\pi}{15}} \langle Y_{l'}^{m'} | Y_2^{-1} | Y_l^m \rangle - \sqrt{\frac{2\pi}{15}} \langle Y_{l'}^{m'} | Y_2^1 | Y_l^m \rangle \right] \end{array} \right\}$$

In contrast to the case of the first pulse, different $m$-states are now coupled to states with $m \pm 1$ and $m \pm 2$, thus leading to the generation of more complex angular wave packets. By solving the above equations on the interval $0 \leq \tau \leq 1$, one is able to decompose the resulting wavepackets in the basis of rotor eigenstates, and to define their time-dependent dynamics.

As mentioned in the main text, the induced unidirectional rotation is accompanied by the confinement of the molecular angular distribution to the plane defined by polarization vectors of the two pulses. We introduce an observable $\langle \cos^2 \varphi \rangle$ (where $\varphi$ is the azimuthal angle) correlated with molecular confinement to the plane. One may present the quantum-mechanical average as $\langle \psi | \cos^2 \varphi | \psi \rangle = \frac{1}{2} + \frac{1}{4}\left(\langle \psi | e^{+i2\varphi} | \psi \rangle + c.c\right)$. Using the expansion of the wavefunction $\psi$ in spherical harmonics $\psi = \sum_{l=0}^{\infty} \sum_{m=-l}^{m=+l} C_{l,m} Y_l^m$ and the explicit expression for $Y_l^m$

$$Y_l^m(\theta, \varphi) = \sqrt{\frac{(2l+1)(l-m)!}{4\pi(l+m)!}} P_l^m(\cos\theta) e^{im\varphi} \quad (12)$$

(where $P_l^m(\cos\theta)$ is the associated Legendre polynomial), one arrives at

$$\langle \psi | e^{+i2\varphi} | \psi \rangle = \sum_{l=0}^{\infty} \sum_{m=-l}^{l} \sum_{l'=0}^{\infty} \sum_{m'=-l'}^{l'} C_{l,m}^* C_{l',m'} \langle Y_l^m | e^{+i2\varphi} | Y_{l'}^{m'} \rangle =$$
$$= \frac{1}{2} \sum_{l'=0}^{\infty} \sum_{l=0}^{\infty} \sum_{m=-l}^{l} C_{l,m}^* C_{l,m-2} \sqrt{\frac{(2l+1)(2l'+1)(l-m)!(l-m+2)!}{(l+m)!(l+m-2)!}} \int_0^\pi P_l^m(\cos\theta) P_{l'}^{m-2}(\cos\theta) \sin\theta d\theta$$
$$(13)$$

Here we used the fact that $e^{+i2\varphi}$ couples only $m$ and $m-2$ eigenstates.

For the overlap integral of associated Legendre polynomials in (13) we used Wong's formula[30]:

$$\int_0^\pi P_{l_1}^{m_1}(\cos\theta) P_{l_2}^{m_2}(\cos\theta) \sin\theta d\theta = \sum_{p_1=0}^{p_{1\max}} \sum_{p_2=0}^{p_{2\max}} a_{l_1,m_1}^{p_1} a_{l_2,m_2}^{p_2} \times$$
$$\times \frac{\Gamma(\frac{1}{2}(l_1+l_2-m_1-m_2-2p_1-2p_2+1))\Gamma(\frac{1}{2}(m_1+m_2+2p_1+2p_2+2))}{\Gamma(\frac{1}{2}(l_1+l_2+3))} \quad (14)$$

where

$$a_{l,m}^{p} = \frac{(-1)^{p}(l+m)!}{2^{m+2p}(m+p)!p!(l-m-2p)!}$$

$p_{max} = [(l-m)/2]$, which is the integer part of $(l-m)/2$ (15)

$l = 0,1,2,....$ and $0 \leq m \leq l$

In order to calculate the overlap integral for negative $m$'s, we used the formula:

$$P_{l}^{-m}(\cos\theta) = (-1)^{m}\frac{(l-m)!}{(l+m)!}P_{l}^{m}(\cos\theta) \qquad (16)$$

The time-dependent thermally averaged value of the azimutal confinement factor $\langle\cos^{2}\varphi\rangle$ calculated by this technique is presented in Fig. 8.

# Appendix B: Double-pulse scheme: Finite-difference time-domain simulation

In order to avoid algebraic complications in treating the double-pulse scheme in the basis of free rotor eigenstates, we undertook a direct numerical solution of the problem by Finite-Difference Time-Domain (FDTD) method. After introducing the dimensionless time, $\tau = (\hbar/I)t$ (where $I$ is the moment of inertia of the molecule), the time-dependent Schrödinger equation takes the following form

$$i\frac{\partial \psi}{\partial \tau} = \frac{1}{2}\vec{\ell}^{\,2}\psi + v(\theta,\varphi,\tau)\psi \qquad (17)$$

The angular momentum operator squared is given by

$$\vec{\ell}^{\,2} = -\left(\frac{\partial^2}{\partial \theta^2} + \frac{1}{\tan\theta}\frac{\partial}{\partial \theta} + \frac{1}{\sin^2\theta}\frac{\partial^2}{\partial \varphi^2}\right). \qquad (18)$$

(3)

The dimensionless interaction potential describing the interaction of the pulse with the induced polarization is:

$$v(\theta,\phi,\tau) = -\frac{I}{4\hbar^2}\varepsilon^2(\tau)\Delta\alpha\,\cos^2\beta \qquad (19)$$

Compared to (2), we consider pulse polarization vector pointing in an arbitrary direction, and $\beta = \beta(\theta,\varphi)$ is the angle between the molecule axis and this vector. In addition, we omitted the insignificant angle-independent term in (2). For our specific problem, the effect of each of the two laser pulses on the wavefunction was considered in the impulsive approximation (5)

$$\psi^{(+)} = \exp[iP\cos^2\beta]\psi^{(-)}, \qquad (20)$$

where the pulse strength, $P$ is defined in (3). The system evolution between the pulses (and after them) is governed by the free Hamiltonian that is azimutally symmetric. This allows us to reduce the two-dimensional time-dependent problem to a set of one-dimensional ones by performing the Fourier transformation in the azimuthal variable $\varphi$:

$$\psi(\theta,\varphi,\tau) = \sum_{m=-\infty}^{\infty} f_m(\theta,\tau)\frac{\exp(im\varphi)}{2\pi}, \qquad (21)$$

where

$$f_m(\theta,\tau) = \int_0^{2\pi} \psi(\theta,\varphi,\tau)\exp(-im\varphi)d\varphi \tag{22}$$

From Schrödinger equation, we obtain an infinite set of equations

$$\frac{\partial f_m}{\partial \tau} = \frac{i}{2}\left(\frac{\partial^2 f_m}{\partial \theta^2} + \frac{1}{\tan\theta}\frac{\partial f_m}{\partial \theta} - \frac{m^2 f_m}{\sin^2\theta}\right), \quad m = 0, \pm 1, \pm 2, \ldots \tag{23}$$

Notice that there are cases when only *one* equation of the set (23) has to be solved. This is, for example, the case when one of the eigenstates of the free rotor, $Y_l^m(\theta,\varphi)$ is kicked by a pulse polarized along the $z$-axis. In this case, $\beta$ in (20) coincides with the polar angle $\theta$, and the wavefunction just after the pulse becomes

$$\psi^+(\theta,\varphi) = \exp(iP\cos^2\theta)\sqrt{\frac{(2l+1)(l-m)!}{4\pi(l+m)!}}\, P_l^m(\cos\theta)\, e^{im\varphi} =$$
$$= f_m(\theta,0)\frac{\exp(im\varphi)}{2\pi} \tag{24}$$

where $P_l^m$ is the associated Legendre polynomial.

In the general case, we use the Fast Fourier Transform (FFT) algorithm to calculate $f_m(\theta,\tau=0)$.

In order to solve the set of equations (23) numerically, one needs to discretize it on a $\theta$-grid. A natural choice for the grid points is $[0, \delta\theta, 2\delta\theta, \ldots, \pi]$ (here $\delta\theta$ is the grid step), but then one faces singularities in the coefficients of Eqs (23) at the north pole, $\theta = 0$, and at the south pole, $\theta = \pi$. There are several ways of dealing with this problem, such as imposing the pole conditions[31,32], or using Fourier decomposition[33]. We choose another method[34], because of its simplicity. We shift the grid by half a distance between adjacent grid points, to avoid placing points at the poles: $[\delta\theta/2, 3\delta\theta/2, \ldots, \pi - \delta\theta/2]$. Discretizing Eq. (23) and using the central-difference approximation for the first derivative, we obtain:

$$\frac{df_m^i}{d\tau} = \frac{i}{2}\left(\frac{f_m^{i-1} - 2f_m^i + f_m^{i+1}}{\delta\theta^2} + \frac{f_m^{i+1} - f_m^{i-1}}{2\delta\theta\tan\theta_i} - \frac{m^2 f_m^i}{\sin^2\theta_i}\right), \tag{25}$$

where the superscript $i$ numerates the points on the grid. For the end-points of the grid, the values of the function $f_m(\theta)$ at "non-physical" arguments $-\delta\theta/2$ and $\pi + \delta\theta/2$ are needed to calculate the derivatives. Using the symmetry of the wavefunction in spherical coordinates

$$\psi(-\theta,\varphi) = \psi(\theta,\varphi+\pi), \quad \psi(\pi+\theta,\varphi) = \psi(\pi-\theta,\varphi+\pi), \tag{26}$$

we obtain the following boundary conditions for $f_m(\theta,\tau)$ at these "non-physical" points:

$$f_m(-\delta\theta/2) = (-1)^m f_m(\delta\theta/2) \, ; \, f_m(\pi+\delta\theta/2) = (-1)^m f_m(\pi-\delta\theta/2) \qquad (27)$$

To propagate the wavefunction in time, we used the Crank-Nicolson method[35]

$$f_m(\tau+\delta\tau) = Uf_m(\tau) = \frac{1-i\mathsf{H}\delta\tau/2}{1+i\mathsf{H}\delta\tau/2} f_m(\tau), \qquad (28)$$

which is second order accurate in $\delta\tau$, and is explicitly unitary ($U^\dagger = U^{-1}$). Here matrix $\mathsf{H}$ presents the discretized Hamiltonian corresponding to the set of equations (25). In order to avoid heavy matrix inversion and multiplication, we rewrite (28) as:

$$f_m(\tau+\delta\tau) = \left(\frac{2}{1+i\mathsf{H}\delta\tau/2} - 1\right) f_m(\tau) = 2\chi(\tau) - f_m(\tau), \qquad (29)$$

where $\chi(\tau)$ is the solution of a tridiagonal linear system

$$(1+i\mathsf{H}\delta\tau/2)\chi(\tau) = f_m(\tau) \qquad (30)$$

Solving the tridiagonal system (30) by backward/forward sweep method is much more efficient numerically than a direct implementation of Eq.(28).

After calculating numerically $f_m(\theta,\tau)$ for all $m$, we use (21) to determine the time-dependent wavefunction $\psi(\theta,\varphi,\tau)$.

For a finite temperature system, this procedure was repeated for all relevant initial $|\ell,m\rangle$ states (which are populated according to the thermal distribution) and the corresponding wavefuctions $\psi^{lm}(\theta,\varphi,\tau)$ were defined. The population of the initial $|\ell,m\rangle$ state is given by $W_{\ell m} = \exp(-E_\ell/kT)/Z$, where $E_\ell = hB\ell(\ell+1)$ is the energy of the rigid rotor eigenstate, $B$ is the rotational constant, $Z$ is the partition function, $T$ is the temperature, and $k$ is the Boltzmann constant. The time dynamics of the average value of any observable $\hat{B}$ depending on $\theta$ and $\varphi$, may be calculated according to:

$$\langle\hat{B}\rangle(\tau) = \int_0^{2\pi}\int_0^\pi \left(\sum_{\ell,m} W_{\ell m} |\psi^{lm}(\theta,\varphi,\tau)|^2\right) B(\theta,\varphi)\sin\theta d\theta d\varphi. \qquad (31)$$

In particular, we used this procedure, to calculate the time-dependent azimutal factor $\langle\cos^2\varphi\rangle$ shown in Fig. 8.

[1] M. A. Duguay and J. W. Hansen. Appl. Phys. Lett. **15**, 192 (1969).
[2] C. H. Lin, J. P. Heritage, and T. K. Gustafson. Appl. Phys. Lett. **19**, 397 (1971).
[3] C. H. Lin, J. P. Heritage, and T. K. Gustafson. Phys. Rev. Lett. **34**, 1299 (1975).
[4] For recent reviews on laser-induced alignment, see H. Stapelfeldt, T. Seideman. Rev. Mod. Phys. **75** (2003) 543; T. Seideman and E. Hamilton, Adv. At. Mol. Opt. Phys. **52**, 289, (2006).
[5] T. Seideman. Chem. Phys. **103**, 7887 (1995).
[6] J. Ortigoso, M. Rodriguez, M. Gupta, and B. Friedrich. J. Chem. Phys. **110**, 3870 (1999).
[7] F. Rosca-Pruna and M. J. J. Vrakking, Phys. Rev. Lett. **87**, 153902 (2001).
[8] I. Sh. Averbukh and R. Arvieu, Phys. Rev. Lett. **87**, 163601 (2001);
M. Leibscher, I. Sh. Averbukh, and H. Rabitz, Phys. Rev. Lett. **90**, 213001 (2003);
M. Leibscher, I. Sh. Averbukh, and H. Rabitz, Phys. Rev. A **69**, 013402 (2004).
[9] M. Renard, E. Hertz, S. Guérin, H. R. Jauslin, B. Lavorel, and O. Faucher, Phys. Rev. A **72**, 025401 (2005).
[10] C. Z. Bisgaard, M. D. Poulsen, E. Péronne, S. S. Viftrup, and H. Stapelfeldt, Phys. Rev. Lett. **92**, 173004 (2004).
[11] K. F. Lee, I. V. Litvinyuk, P. W. Dooley, M. Spanner, D. M. Villeneuve, and P. B. Corkum, J.Phys. B: At., Mol., Opt. Phys. **37**, L43 (2004).
[12] C. Z. Bisgaard, S. S. Viftrup, and H. Stapelfeldt, Phys. Rev. A **73**, 053410, (2006).
[13] D. Pinkham and R.R. Jones, Phys. Rev. A **72**, 023418 (2005).
[14] J. Karczmarek, J.Wright, P.Corkum, and M. Ivanov, Phys. Rev. Lett. **82**, 3420 (1999).
[15] I.V. Litvinyuk, Kevin F. Lee, P.W. Dooley, D.M. Rayner, D.M. Villeneuve, and P.B. Corkum. Phys. Rev. Lett. **90**, 233003 (2003).
[16] S. Fleischer, I. Sh. Averbukh and Y. Prior, PRA **74** 041403 (2006).
[17] S. Fleischer, I. Sh. Averbukh and Y. Prior, PRL **99** 093002 (2007).
[18] R.A.Bartels, T.C.Weinacht, N.Wagner, M. Baertschy, Chris H. Greene, M.M. Murnane, and H.C. Kapteyn, Phys. Rev. Lett **88**, 013903 (2002).
[19] V. Kalosha, M. Spanner, J. Herrmann, and M. Ivanov, Phys. Rev. Lett., **88**, 103901 (2002).
[20] R. Velotta, N. Hay, M. B. Mason, M. Castillejo, J. P. Marangos, Phys. Rev. Lett. **87**, 183901 (2001).
[21] M. Kaku, K. Masuda, and K. Miyazaki. Jpn. J. Appl. Phys. **43**, L591 (2004).
[22] J. Itatani, D. Zeidler, J. Levesque, M. Spanner, D.M. Villeneuve, P.B. Corkum, Phys. Rev. Lett. **94,** 123902 (2005).
[23] E. J.Brown, Qingguo Zhang and M. Dantus. J.Chem Phys **110**, 5772 (1999).
[24] M. Comstock, V. Senekerimyan, and M. Dantus. J. Phys. Chem. A **107**, 8271 (2003).
[25] J. Itatani et al. Nature **432**, 867 (2004).
[26] B. K. McFarland, J. P. Farrell, P. H. Bucksbaum, M. Gühr, *Science* **322**, 1232 (2008)
[27] M. W. Wefers, H. Kawashima and K. A. Nelson, J.Phys. Chem. Solids, **57**, 1425 (1996).
[28] Y. Khodorkovsky, I. Sh. Averbukh (to be published)
[29] M. Leibscher, I. Sh. Averbukh, P. Rozmej, and R. Arvieu, Phys. Rev. A **69**, 032102 (2004)
[30] B.R. Wong, J. Phys. A: Math Gen **31**, 1101 (1998)

[31] S. Y. K. Yee, Monthly Weather Review **109**, 501 (1981).

[32] J. Shen, J. Sci. Comput. **20**, 1438 (1999).

[33] C. E. Dateo, V. Engel, R. Almeida and H. Metiu, Comp. Phys. Commun. **63**, 435 (1991); C. E. Dateo and H. Metiu, J. Chem. Phys. **95**, 7392 (1991).

[34] M.-C. Lai and W.-C. Wang, Numer. Meth. Partial Diff. Eq. **18**, 56 (2002); M.-C. Lai, W.-W Lin and W. Wang, J. Numer. Anal. **22**, 537 (2002).

[35] W. H. Press, S. A. Teukolsky, W. T. Vetterling and B. P. Flannery, *Numerical Recipes in Fortran 77: The Art of Scientific Computing* (2001).